# Telecom wavelength single quantum dots with very small excitonic fine-structure splitting

Andrei Kors, Johann Peter Reithmaier, and Mohamed Benyoucef[*]

Institute of Nanostructure Technologies and Analytics (INA), Center for Interdisciplinary Nanostructure Science and Technology (CINSaT), University of Kassel, Heinrich-Plett-Str. 40, 34132 Kassel, Germany

We report on molecular beam epitaxy growth of symmetric InAs/InP quantum dots (QDs) emitting at telecom C-band (1.55 µm) with ultra-small excitonic fine-structure splitting of ~2 µeV. The QDs are grown on distributed Bragg reflector and systematically characterized by micro-photoluminescence (µ-PL) measurements. One order of magnitude of QD PL intensity enhancement is observed in comparison with as-grown samples. Combination of power-dependent and polarization-resolved measurements reveal background-free exciton, biexciton and dark exciton emission with resolution-limited linewidth below 35 µeV and biexciton binding energy of ~1 meV. The results are confirmed by statistical measurements of about 20 QDs.

---

[*] Corresponding author. email: m.benyoucef@physik.uni-kassel.de

Various quantum photonics potential applications such as future quantum information processing[1], communication and cryptography[2] require entangled and indistinguishable photons. Single semiconductor quantum dot (QD) as a source of quantum light[3,4] is currently under active investigation particularly due to deterministic nature of produced photons and possibilities of integration in well-developed semiconductor technology. Polarization-entangled photon pairs are produced during biexciton–exciton (XX–X) radiative cascade of a single QD[5]. One of the challenges on the way towards ideal entangled photon sources is an asymmetry of the QDs which leads to energetic separation of the bright excitonic states, so-called fine-structure splitting (FSS). When the FSS is larger than the emission linewidth, the fidelity of entangled photons is significantly reduced.

It has been shown that FSS can be reduced by post-growth energy splitting compensation methods[6], however, it would be more practical to realize an approach that allows high quality QD to be formed with smallest possible FSS. The most advanced results so far were demonstrated in the GaAs-based material system by applying special partial capping and annealing technology[7] or droplet etching method[8]. However, for long-distance optical fiber[9] applications it is necessary to have an emission with minimum attenuation at telecom C-band (1.55 μm). There are various methods allowing low QD density and single-photon emission at telecom wavelengths. The most common approach is based on the InP material system. For example, variations of the Stranski-Krastanov (S-K) QD growth process, such as ripening[10,11] and double-capping methods, show high-purity single-photon emission[12]. Another approach is based on droplet epitaxy with low FSS values[13] and it demonstrates electrically injected entangled photon emission[14]. Strategy of QDs growth on $C_{3v}$ symmetric (111) substrates, leading to vanishing FSS, also should be mentioned[15]. Recently, single[16,17] and polarization-entangled photons[18] could be also shown in the GaAs material system under certain conditions.

Our approach towards low-density high-symmetry QDs emitting at telecom wavelength is based on a ripening technique[10,11,19]. In our previous works on InAs, QDs were embedded in AlGaInAs matrix to avoid complex arsenic-phosphorus reactions. This approach resulted in a single-photon emission and relatively large FSS of 20 μeV[11]. However, according to theory[20], smaller intrinsic FSS values are predicted for QDs embedded in InP matrix. Recently, we reported on a telecom single InAs/InP QD emission embedded in an InP-based photonic crystal microcavity with a FSS value down to 5 μeV[19].

Here we report on systematic studies of single InAs/InP QDs emitting at 1.55 μm. QDs are characterized by atomic force microscopy (AFM), macro- and micro- photoluminescence (μ-PL). For enhanced PL emission, QDs are grown on top of a distributed Bragg reflector (DBR) which itself is characterized by reflectivity measurements. Finally, large amount of single QDs are investigated by combination of power- and polarization-depended μ-PL measurements. Both as-grown and DBR samples reveal the existence of a biexciton (XX) - exciton (X) system with low background signal and highly reduced excitonic FSS values down to 2 μeV.

The investigated samples are grown by solid-source molecular beam epitaxy (MBE) on (100) oriented InP substrates. The QDs are grown by depositing nominally 2 monolayers (MLs) of InAs at a growth temperature of 490 °C. The growth of low-density QDs is based on the ripening process to form larger QDs emitting at the telecom wavelength of 1.55 μm[10,11,19]. For optical investigations the sample was capped with a 100 nm thick InP layer. For morphological characterizations a second QD layer is grown on top of the sample at the same conditions as for the embedded QD layer. In order to enhance PL emission, another QD sample is grown on InP/InAlGaAs distributed Bragg reflectors (DBRs).

The concept of the ripening technique is demonstrated by series of a reflection high energy electron diffraction (RHEED) patterns as shown in Fig.1(c). Left image shows a sample surface reconstruction directly after deposition of a 2 MLs thick InAs layer. One can see a streaky pattern, indicating an atomically flat surface. During a cooling of the sample a gradual transition from streaky to spotty pattern is observed indicating the transition from 2D to 3D structures and the formation of QDs[21]. By controlling the ripening time, samples with a low dot density emitting at telecom wavelength can be realized. Inset of Fig. 1(a) shows a $1 \times 1$ μm$^2$ AFM scan of the as-grown sample with a QD density of $2 \times 10^9$ cm$^{-2}$. Single QDs with different sizes and some nanoclusters are clearly visible in the AFM image. The smaller structures have a round-shape and bigger structures have a square-shaped tendency with a smooth surface in between the DQs. It should be mentioned that the shape and size of embedded QDs might be changed during the arsenic-phosphorus exchange reactions and InP capping layer growth. Corresponding macro-PL spectrum is labeled as QDs in Fig. 1(a). Broad wavelength emission range with a maximum wavelength emission is centered around 1.55 μm indicates a size variation of embedded QDs. Despite of a relatively high excitation power, observed emission is quite weak, which is an indication of a low QD density.

For enhanced PL emission, QDs are embedded in a DBR structure. Fig. 1(b) shows a reflectivity spectrum of the investigated DBR sample. Reflectivity value is reaching 99% with a stop band of around 100 nm centered at telecom C-band and demonstrates a good sample quality. Optical properties of the DBR sample were directly compared with the properties of the as-grown sample, which indicates an enhanced PL emission by an order of magnitude at 1.55 μm as clearly visible in Fig. 1(a). For the μ-PL spectroscopy measurements the sample is mounted in a liquid helium flow cryostat and kept at cryogenic temperatures around 5 K. The samples are excited with a continuous wave laser (532 nm), which is focused by a microscope objective (NA = 0.7) to a spot size of ~1 μm. Emitted light is spectrally filtered by a 0.75 m focal length spectrometer equipped with a liquid nitrogen cooled InGaAs array detector.

Figure 2 displays typical µ-PL spectra of single QDs from DBR and as-grown samples recorded at 180 nW excitation power (QD 1, QD 2) and at 2850 nW (QD3), respectively. Due to relatively low QD densities emitting at telecom C-band wavelengths, $\mu$-PL measurements are performed without the need for mesa structures. PL spectra exhibits background-free quantum-dot emission with exciton (X) and biexciton (XX) transitions around 1.55 µm. Note that the XX transition is observed below the X energy for all investigated QDs. Emission linewidth is determined by the Gaussian fitting of the recorded spectra. A full width at half maximum (FWHM) value is below our PL setup resolution limit of 35 µeV for the best QDs. The Gaussian-like shape of excitonic lines might be related to PL setup resolution and a spectral broadening due to the charge fluctuations around the QD. Spectral linewidth could be further reduced by using a high resolution PL setup and optimizing the QDs growth conditions. Insets of Fig. 2(b) show 2D and 3D AFM images of a nearly round-shaped single QD from the as-grown sample. Corresponding µ-PL spectrum from this sample is shown in Fig.2(c). It has similar properties with QDs grown on the DBR structure. Therefore, such as-grown sample could be used for direct probing of single QDs properties and optimized conditions could be later implemented into cavity structures. By performing the power-depended measurements of QD2 and QD3 (Fig. 2(d)), an order of magnitude of PL intensity enhancement of QDs emission grown on DBR structure compared with the QDs in bulk is obtained. This value is in agreement with the macro-PL results (Fig. 1(a)).

Figure 3(a) shows four PL spectra of the single QD taken at different excitation powers. At relatively low excitation power of 20 nW the X line dominates the spectrum and XX line starts to appear with a resolution-limited FWHM below 35 µeV as it is shown in Fig. 3(c). At 138 nW the XX linewidth slightly broadens and reaches a value of 54 μeV, whereas the X linewidth value stays nearly constant around 45 μeV even at higher excitation powers. At 85 nW an additional resolution-limited line appears between X and XX and increases approximately linearly with the excitation power. Fig. 3(b) shows integrated PL intensities of the X and XX transitions as a function of laser excitation power. The dashed lines are linear fits to the data in a double-logarithmic plot, showing almost ideal linear and quadratic behaviors for the X and XX lines, respectively. When X line reaches the saturation region around 100 nW, additional excitonic lines start to appear at lower energies and are clearly resolved at 138 nW. Signatures of similar lines are also observed for different QDs and can be seen in Fig. 2(b, c). However, detailed investigation of this phenomenon is beyond the scope of the current paper.

Photoluminescence intensity mapping performed on the same QD described in Fig. 3(a) is recorded at 300 nW for different linear polarization angles and illustrated in Fig. 4(a) in a logarithmic scale. The XX line shows more pronounced energy shift of the peak position. However, due to the resolution limit of the setup, only certain asymmetry of this line is visible in the spectrum. The inset of Fig. 4(a) shows a relative change of the X peak energy as a function of linear polarization angles. Data points and error bars were extracted from Gaussian fits of recorded spectra. Sinusoidal behavior with amplitude of 2 μeV confirms that this line is not originating from a charged exciton, which is spin degenerate according to the Kramers theorem. Clear anti-correlated oscillatory behavior of both lines is confirming the above mentioned signatures of the exciton-biexciton system. The energy difference between X and XX lines $E_X - E_{XX}$ gives a biexciton binding energy value of 1 meV. X and XX lines have a polarization degree $P \equiv \frac{I_{max} - I_{min}}{I_{max} + I_{min}}$ of 0.39 and 0.52, respectively. A characteristic feature of the investigated QD emission lines is the presence of the extra line between X and XX. It has an orthogonal polarization degree of 0.83 determined from the sinusoidal fitting.

The line position is polarization insensitive and therefore we attribute this to a dark excitonic signature[22], however, a detailed dark excitonic analysis is beyond the scope of this paper. For more precise fitting and analysis, we will focus our attention on the X line.

The statistical distribution of the FSS of the X line for about 20 QDs is summarized in Fig. 4(c) exhibiting a small excitonic fine structure splitting down to ~2 μeV. Note that the measured FSS values have a narrow distribution (inset of Fig.4(c)) in agreement with the theoretically predicted values[20] and smaller than previously reported FSS data for InAs/InP based system[13,15]. Furthermore, the biexciton binding energy values are nearly constant and all observed QDs have the same configuration where the X line is located at higher energies of the PL spectrum. This suggests that the structure, composition, and other QD parameters are nearly identical in this spectral range[23]. Biexciton binding energies in these QDs are lower than for earlier reported quantum dashes[24,25] or InAs/InP QDs grown by droplet epitaxy[13] and close to the Stranski-Krastanov InAs/InP QDs[26]. We suppose that the small measured excitonic fine-structure splitting values is a direct consequence of both intrinsic InAs/InP material properties and the ripening technique, which allows the formation of highly symmetric QDs. It should also be noted that during the ripening process QDs are longer exposed to the residual atmosphere in the growth chamber. Therefore QDs morphology, composition and properties could be affected. For example, it was shown that nitrogen incorporation to InAs QDs strongly affects the FSS[27,28]. However, for more detailed analysis additional measurements are required.

In conclusion, we have demonstrated a background-free single QD emission at telecom wavelengths with ultra-small excitonic fine-structure splitting of ~2 μeV. μ-PL measurements reveal resolution-limited excitonic linewidths below 35 μeV. X and XX emission lines were assigned by power and polarization dependent measurements. Typical biexciton binding energies are around 1 meV. These results show the potential of our QDs to generate high-quality polarization-entangled photon pairs on demand at telecom wavelength, suitable for long-distance quantum information applications.

This work was financially supported by the BMBF project German Federal Ministry of Education and Research (BMBF) under the Program "Quantum Communication" (Grant Q.com-H 16KIS0112).

**Figures captions:**

Fig. 1. (a) Photoluminescence spectra of as-grown low-density QD sample (red) and QDs grown in the same conditions on the DBR structure (black). Inset: $1 \times 1$ µm$^2$ AFM image of the as-grown sample. (b) Reflectivity spectrum of a DBR sample. (c) From left to right: evolution in time of RHEED patterns during the QD ripening process.

Fig. 2. (a-b) µ-PL spectra from two different single QDs on DBR structure showing a biexciton (XX) – exciton (X) emission with a resolution-limited excitonic linewidth emitting at telecom C-band. Insets show 2D and 3D AFM images of a nearly round-shaped single QD. (c) Single QD emission from an as-grown sample. (d) Power dependent measurements of excitonic lines from as-grown (round) and DBR (square) QDs.

Fig. 3. (a) $\mu$-PL spectra from a single QD under cw non-resonant excitation, recorded for four different laser excitation powers. (b) Integrated PL intensities of exciton and biexciton as a function of the excitation powers. (c) Resolution limited PL spectrum of the biexciton (XX) line taken at relatively low excitation power of 20 nW.

Fig. 4. (a) Polarization-dependent PL intensity map in a logarithmic scale of a single QD. Inset: The emission energies of X as a function of the linear polarization angles showing an average very small excitonic FSS value of ~2 µeV. (b) X and XX transitions recorded at 0° (red) and 90° (blue) polarization angles and the subtracted spectrum (black), showing vanishing fine-structure splitting from the single QD. (c) X FSS values as a function of exciton emission energy for different QDs. Inset: Statistical distribution of the X FSS. (d) Biexciton binding energies as a function of X emission energy for different QDs. Inset: statistical distribution of the biexciton binding energies.

Figure 1

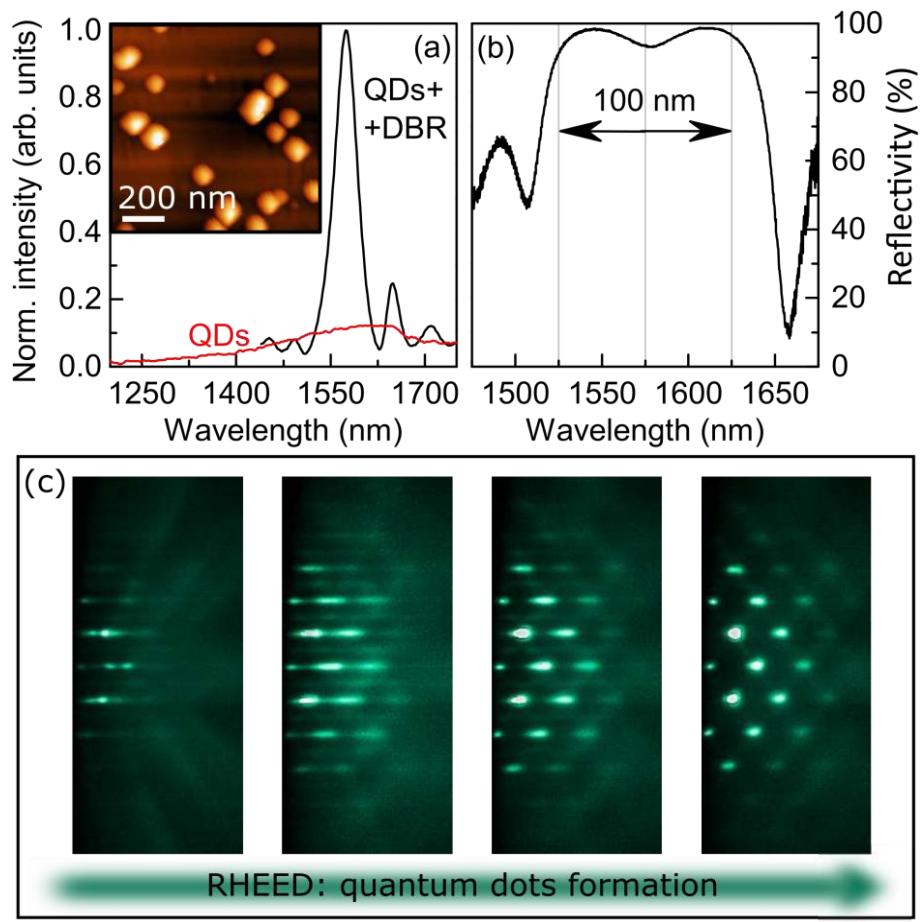

Figure 2

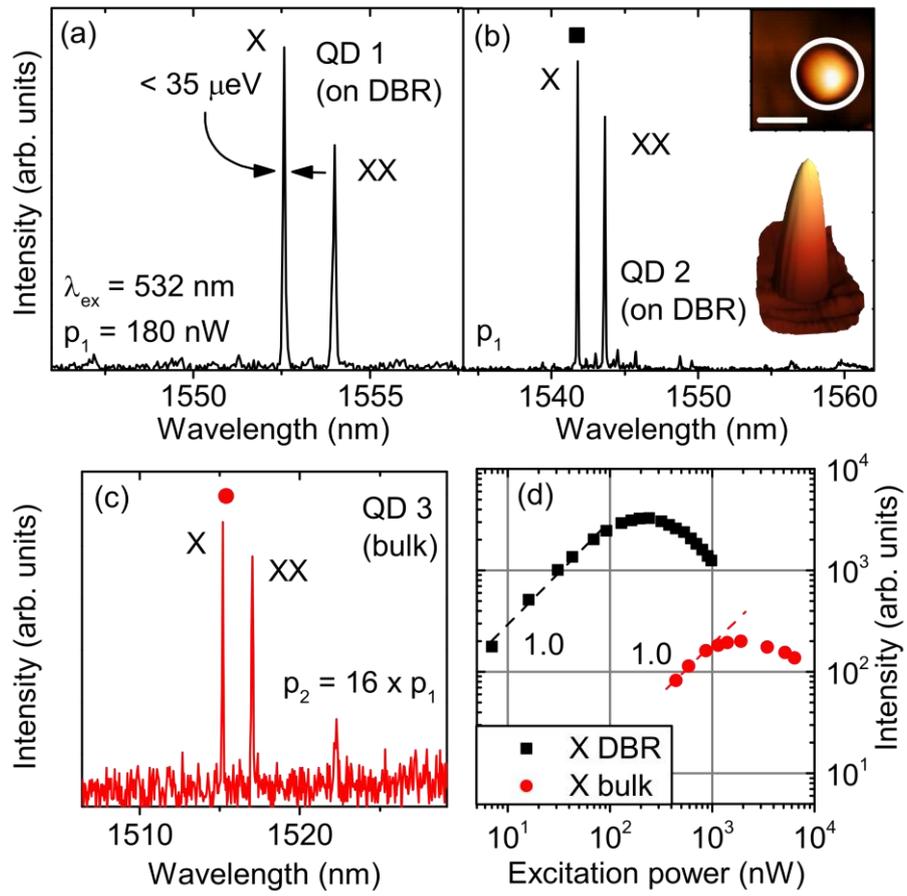

Figure 3

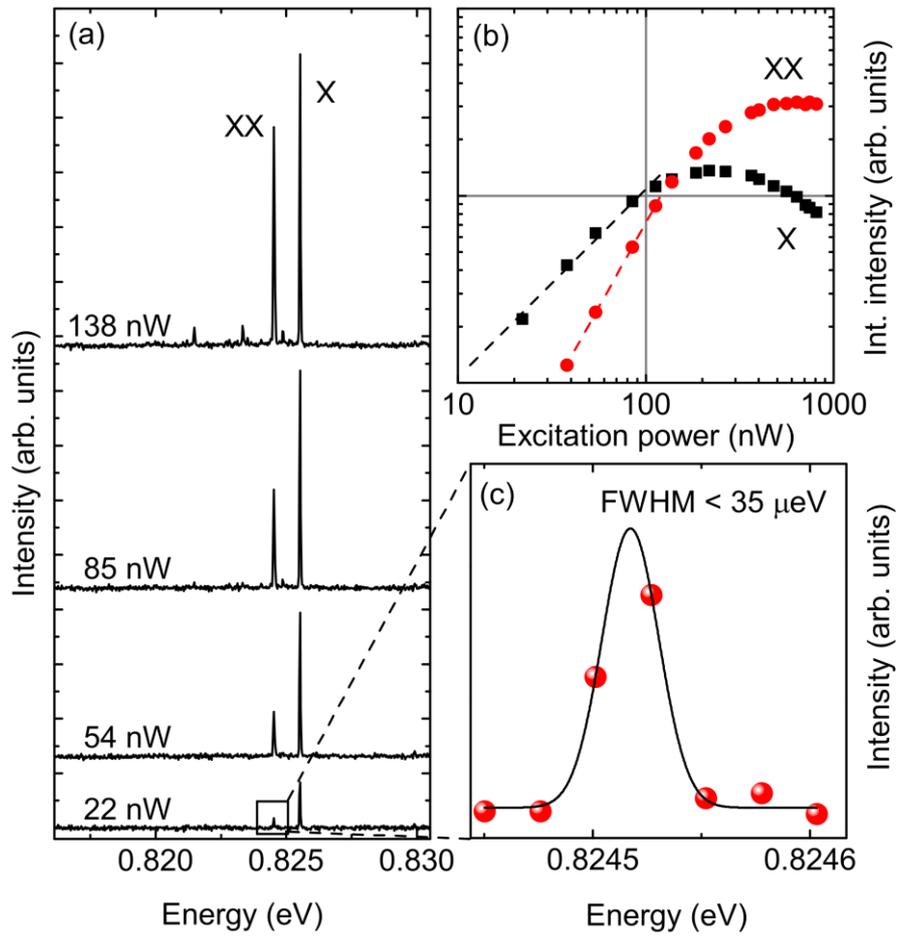

Figure 4

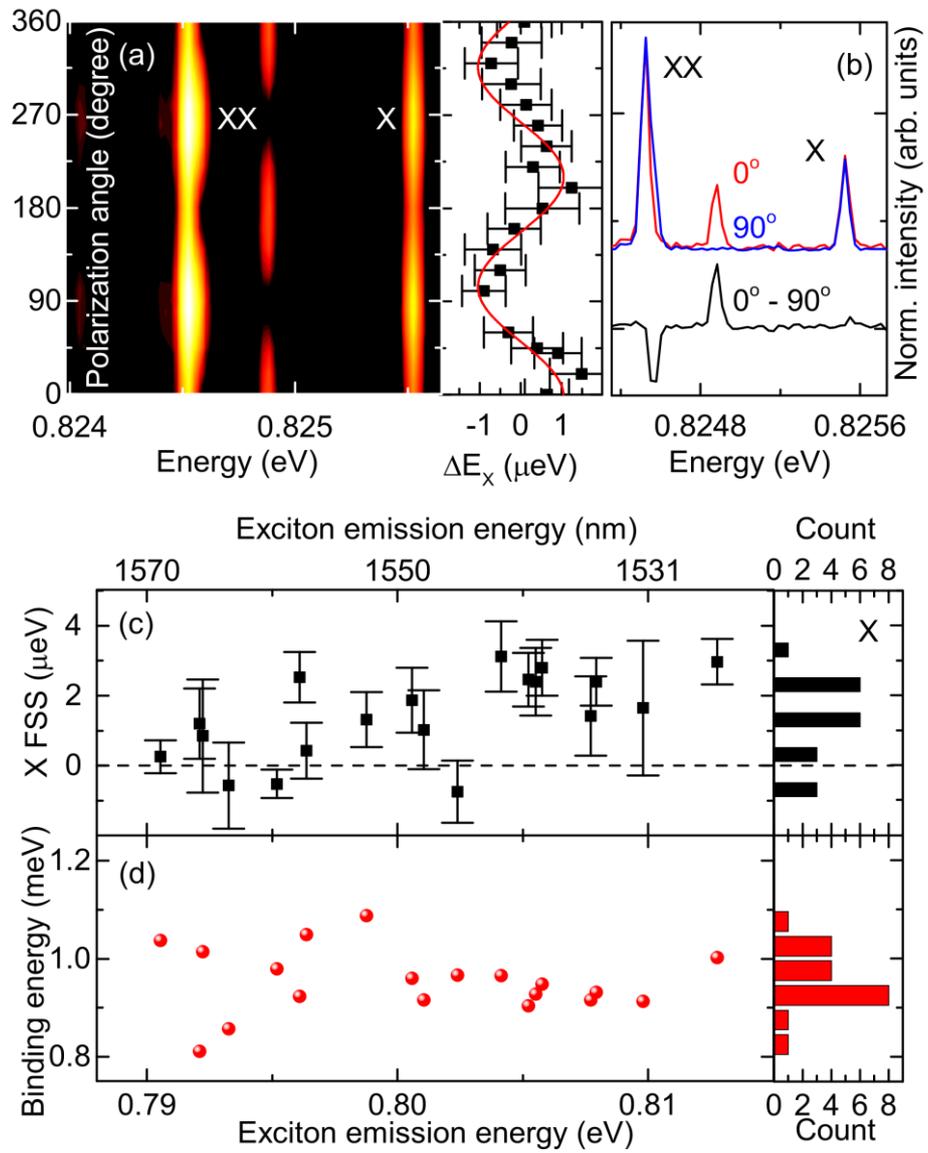